\documentclass[runningheads]{llncs}
\usepackage{graphicx}
\usepackage{amssymb}
\usepackage{mwe}
\usepackage{rotating}
\usepackage{amsmath}
\usepackage{ctable}
\usepackage{float}
\usepackage{bm}

\newcommand{\RNum}[1]{\uppercase\expandafter{\romannumeral #1\relax}}

\usepackage{color}

\begin{document}
\title{Retinal OCT Denoising with Pseudo-Multimodal Fusion Network}
\author{Dewei Hu\inst{1}\and
Joseph D. Malone\inst{2}\and
Yigit Atay\inst{1}\and
Yuankai K. Tao\inst{2}\and
Ipek Oguz\inst{1}}
\authorrunning{D. Hu et al.}
\institute{Vanderbilt University, Dept. of Electrical Engineering and Computer Science, Nashville, TN, USA \and
Vanderbilt University, Dept. of Biomedical Engineering, Nashville, TN, USA}
\titlerunning{OCT Denoising with PMFN}

\maketitle      
\begin{abstract}
Optical coherence tomography (OCT) is a prevalent imaging technique for retina. However, it is affected by multiplicative speckle noise that can degrade the visibility of essential anatomical structures, including blood vessels and tissue layers. Although averaging repeated B-scan frames can significantly improve the signal-to-noise-ratio (SNR), this requires longer acquisition time, which can introduce motion artifacts and cause discomfort to patients. In this study, we propose a learning-based method that exploits information from the single-frame noisy B-scan and a pseudo-modality that is created with the aid of the self-fusion method. The pseudo-modality provides good SNR for layers that are barely perceptible in the noisy B-scan but can over-smooth fine features such as small vessels. By using a fusion network, desired features from each modality can be combined, and the weight of their contribution is adjustable. Evaluated by intensity-based and structural metrics, the result shows that our method can effectively suppress the speckle noise and enhance the contrast between retina layers while the overall structure and small blood vessels are preserved. Compared to the single modality network, our method improves the structural similarity with low noise B-scan from $0.559\pm 0.033$ to $0.576\pm 0.031$.

\end{abstract}
\section{Introduction}
Optical coherence tomography (OCT) is a powerful non-invasive ophthalmic imaging tool~\cite{li2017statistical}. The limited light bandwidth of the imaging technique on which OCT is based upon, low-coherence interferometry~\cite{schmitt1999speckle}, gives rise to speckle noise that can significantly degrade the image quality. In clinical practice, the thickness of the retina layers, such as the ganglion cell layer (GCL), inner plexiform layer (IPL) and retinal nerve fiber layer (RNFL), are of interest ~\cite{tatham2017detecting}. Retinal OCTs also reveal the vascular system, which is important for ocular diseases like diabetic retinopathy~\cite{ouyang2015retinal}. The speckle noise in single frame B-scans makes the border of layers unclear so that it is hard to distinguish adjacent layers, such as the GCL and IPL. The noise also produces bright dots and dark holes that can hurt the homogeneity of layers and affect the visibility of the small vessels within them. A proper denoising method is thus paramount for ophthalmic diagnosis.

Acquiring multiple frames at the same anatomical location and averaging these repeated frames is the mainstream technique for OCT denoising. The more repeated frames are acquired, the closer their mean can be to the ideal ground truth. However, this increases the imaging time linearly, and can cause discomfort to patients as well as increase motion artifacts. Other hardware-based OCT denoising methods including spatial~\cite{avanaki2013spatial} and angular averaging~\cite{schmitt1997array} will similarly prolong the acquisition process. Ideally, an image post-processing algorithm that applies to a single frame B-scan is preferable. Throughout the paper, we denote single frame B-scan as high noise (HN) and frame-average image as low noise (LN).

\begin{figure}[t]
    \begin{tabular}{ccccc}
    \centering
    &High-noise (HN)  & Low-noise (LN)  & Self-fusion of HN & Self-fusion of LN \\
    \rotatebox{90}{\hspace{1cm}ONH} & 
    \includegraphics[width=0.23\linewidth]{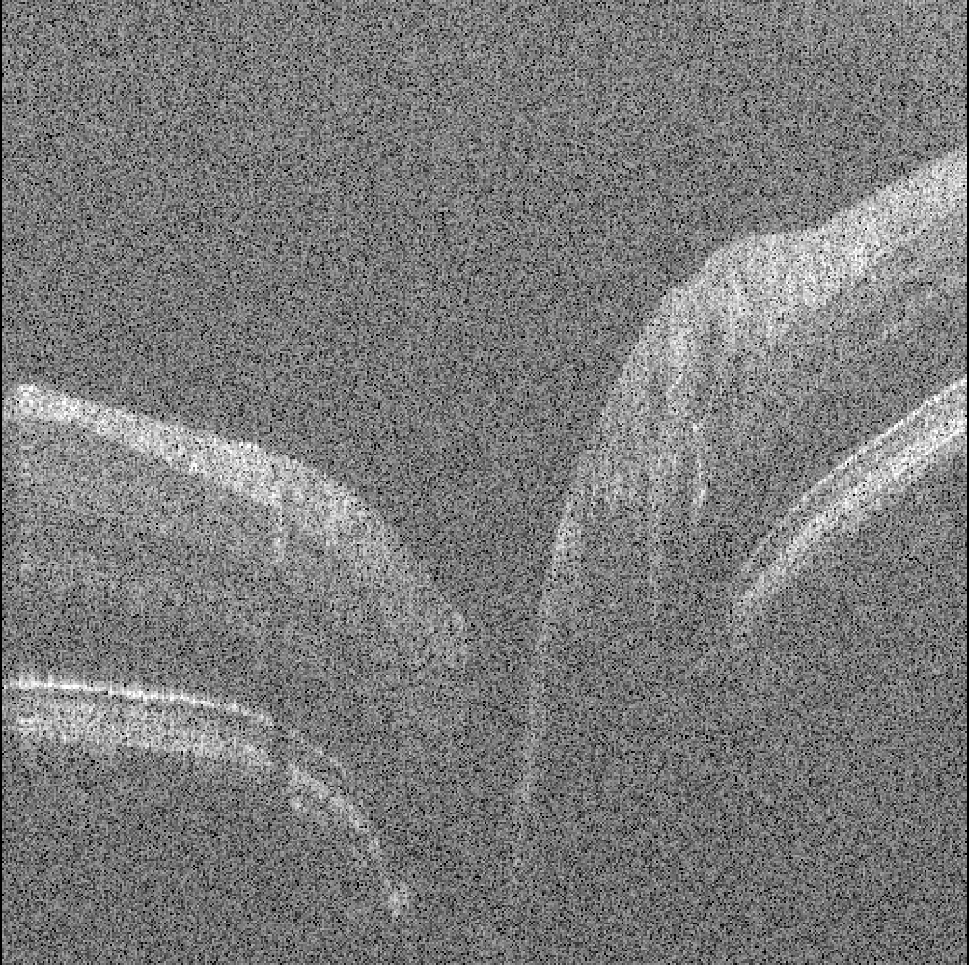}&
    \includegraphics[width=0.23\linewidth]{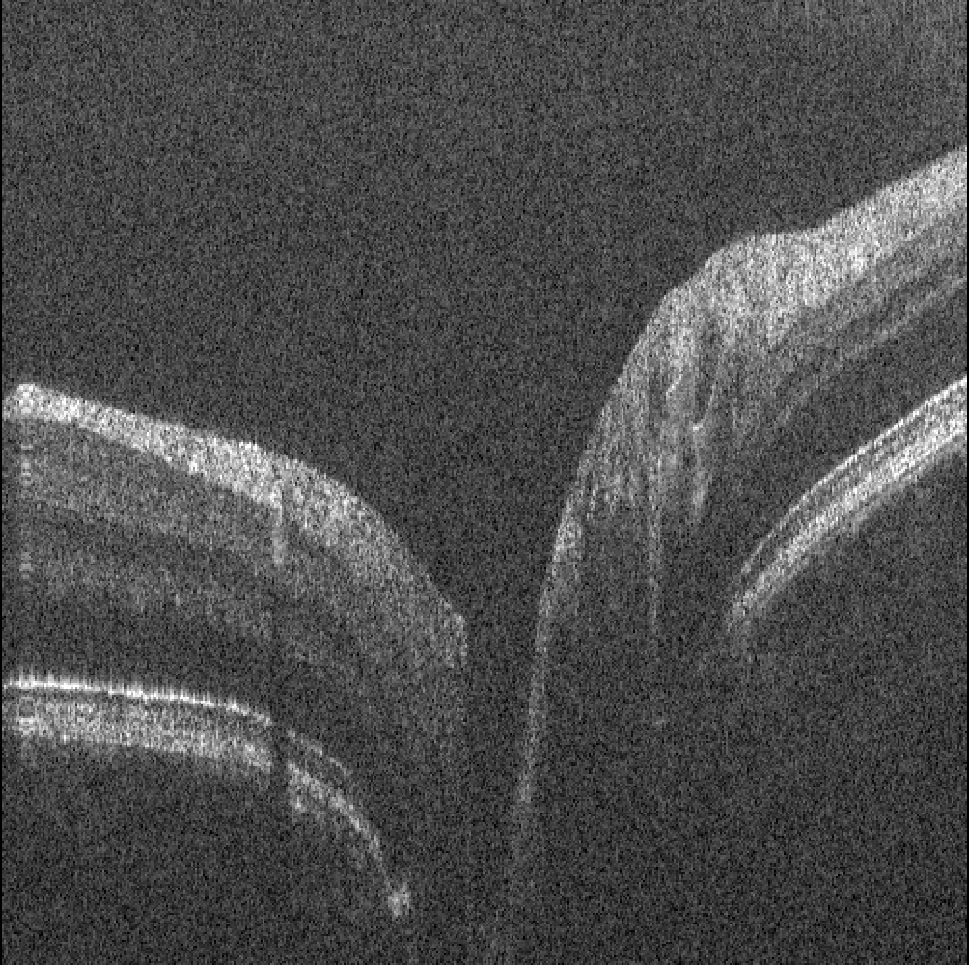}&
    \includegraphics[width=0.23\linewidth]{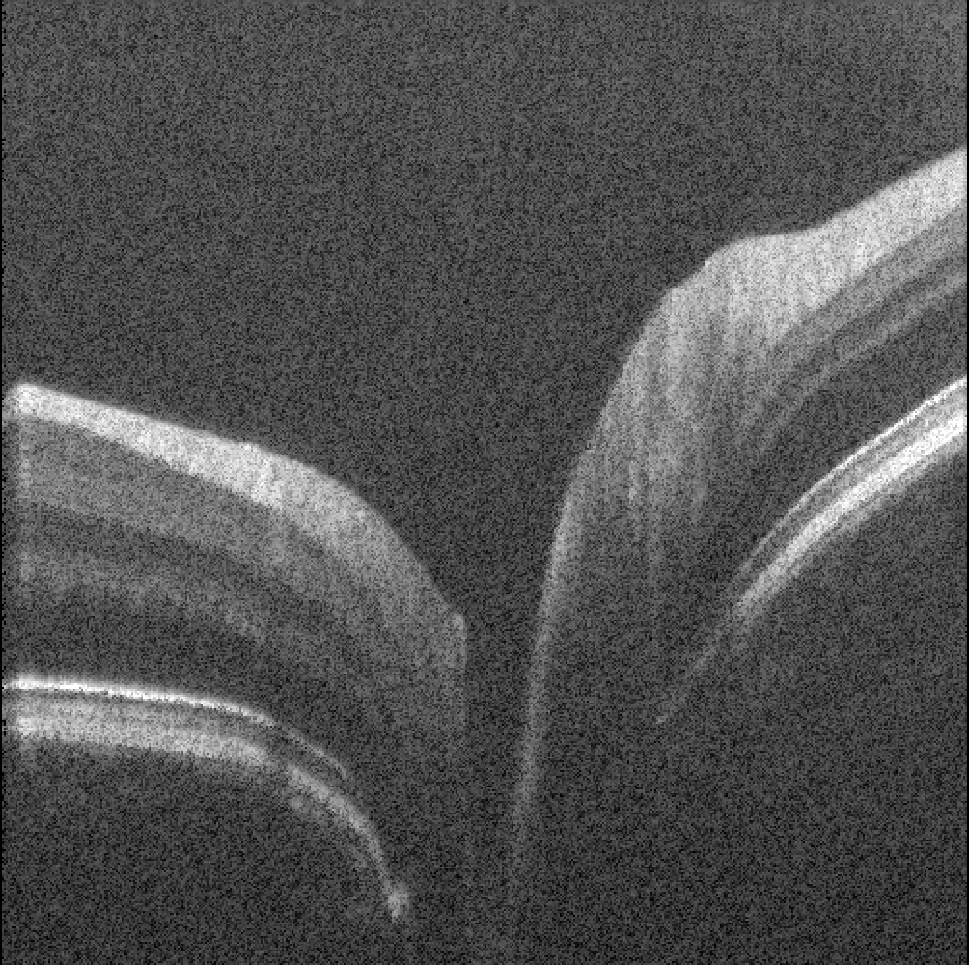}&
    \includegraphics[width=0.23\linewidth]{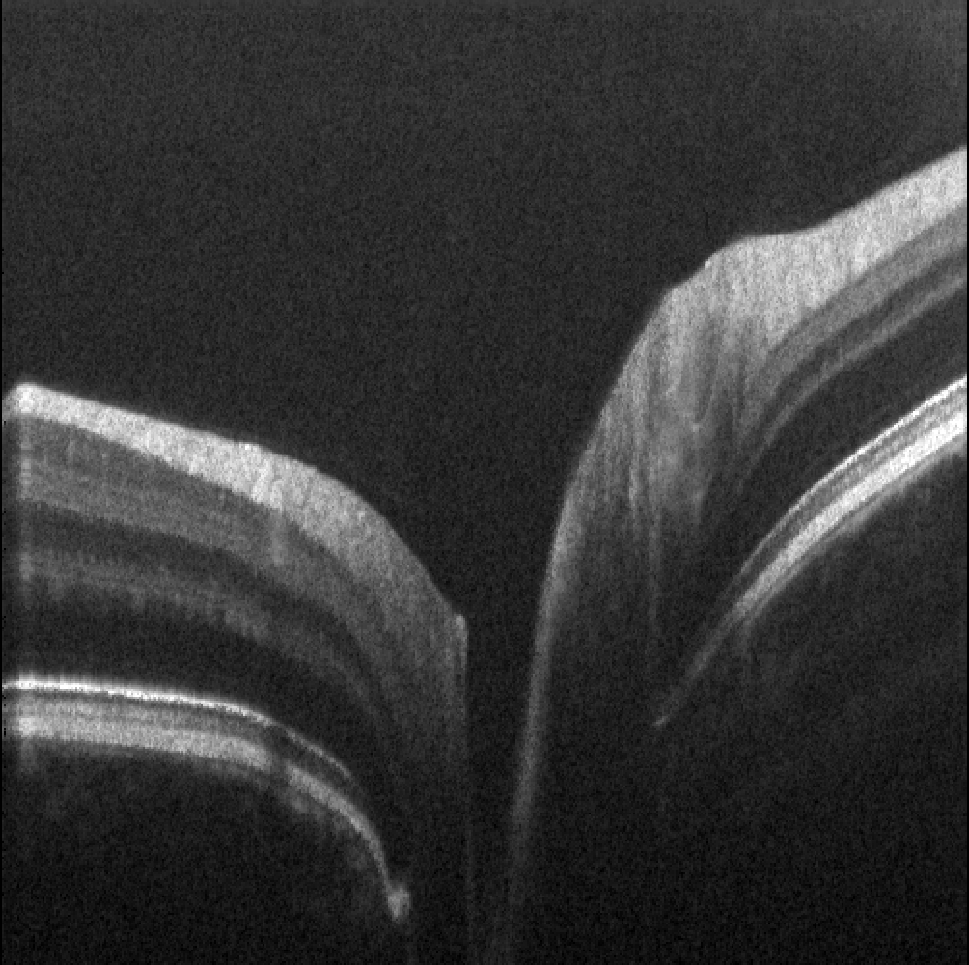}\\
    \rotatebox{90}{\hspace{0.4cm}Fovea} & 
    \includegraphics[width=0.23\linewidth,trim={0cm 4cm 0cm 5cm},clip]{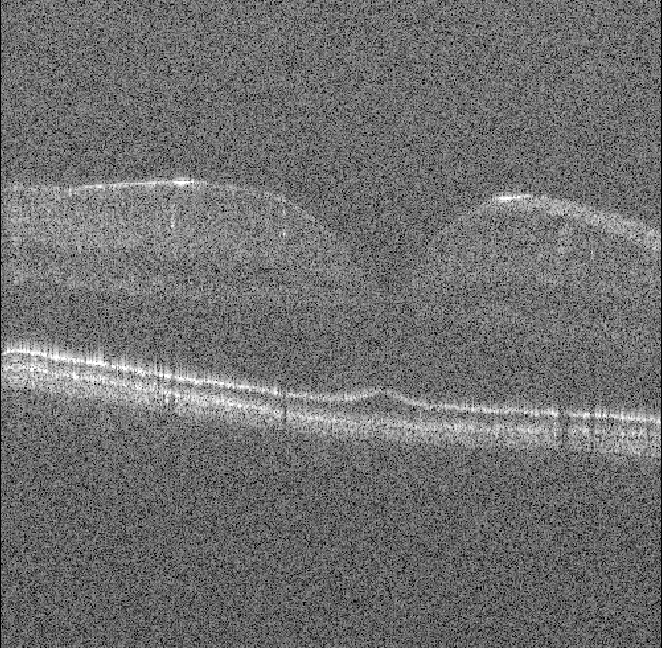}&
    \includegraphics[width=0.23\linewidth,trim={0cm 4cm 0cm 5cm},clip]{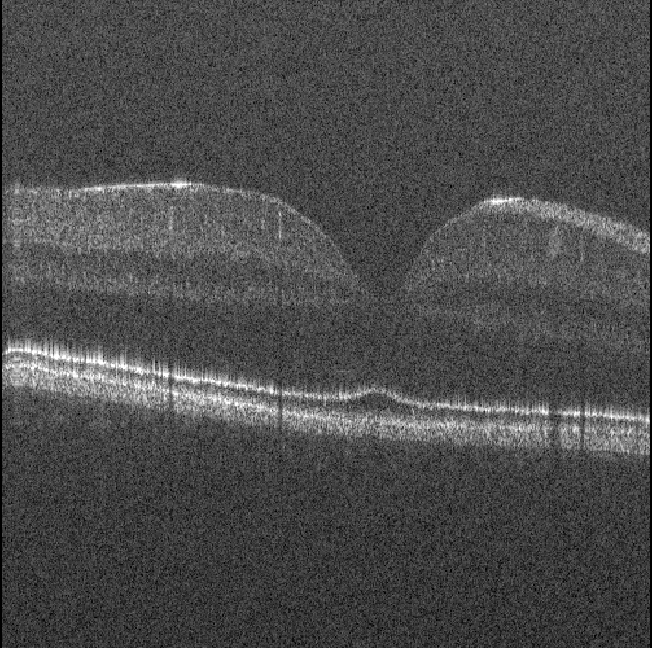}&
    \includegraphics[width=0.23\linewidth,trim={0cm 4cm 0cm 5cm},clip]{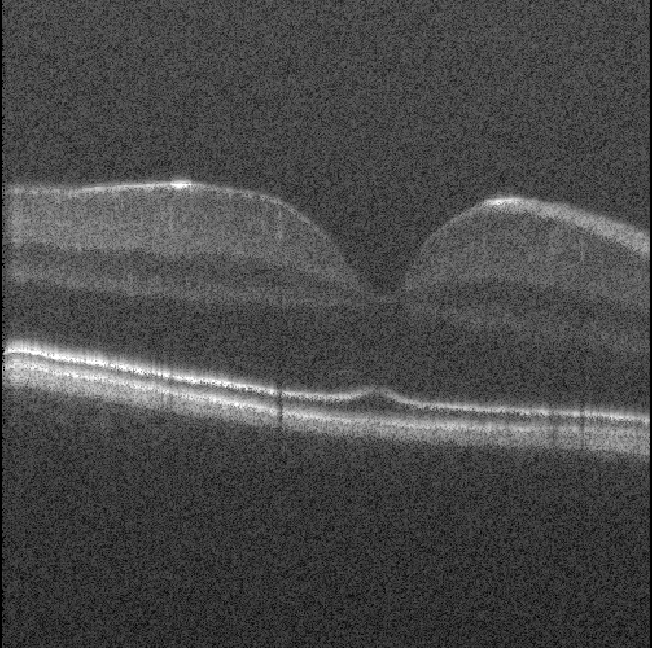}&
    \includegraphics[width=0.23\linewidth,trim={0cm 4cm 0cm 5cm},clip]{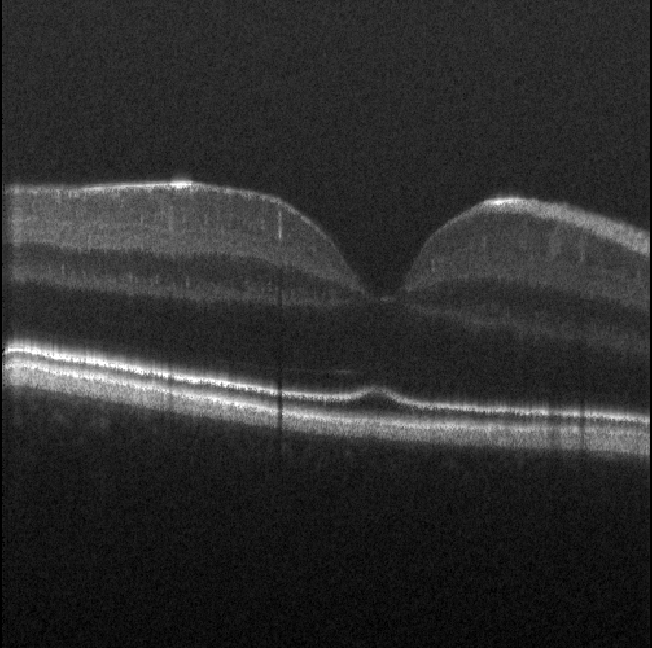}
    \end{tabular}
    \caption{Self-fusion for high-noise (HN) single B-scan and low-noise (LN) 5-average images (excess background trimmed). SNR of the HN images is 101dB.}
    \label{fig:sf}
\end{figure}

The multiplicative nature of speckle noise makes it hard to be statistically modelled, as the variation of noise intensity level in different tissue increases the complexity of the problem~\cite{chen2020dn}. In a recent study, Oguz et al.~\cite{oguz2020self} proposed the self-fusion method for retinal OCT denoising. Inspired by multi-atlas label fusion~\cite{wang2012multi}, self-fusion exploits the similarity between adjacent B-scans. For each B-scan, neighboring slices within radius $r$ are considered as `atlases' and vote for the denoised output. As shown in Fig.~\ref{fig:sf}, self-fusion works particularly well in preserving layers, and in some cases it also offers compensation in vessels. However it suffers from long computation time and loss of fine details, similar to block-matching 3D (BM3D)~\cite{chong2013speckle} and $k$ singular value decomposition (K-SVD)~\cite{kafieh2014three}.  

Deep learning has become the state-of-the-art in many image processing tasks and shown great potential for image noise reduction. Although originally used for semantic segmentation, the U-Net~\cite{ronneberger2015u} architecture enables almost all kinds of image-to-image translation~\cite{isola2017image}. Formulated as the mapping of a high noise image to its `clean' version, the image denoising problem can easily be seen as a supervised learning algorithm. Because of the poor quality of single frame B-scan, more supplementary information and constraints are likely to be beneficial for feature preservation. For instance, observing the layered structure of the retina, Ma et al.~\cite{ma2018speckle} introduce an edge loss function to preserve the prevailing horizontal edges. Devalla et al.~\cite{devalla2019deep} investigate a variation to U-Net architecture so that the edge information is enhanced.

In this study, we propose a novel despeckling pipeline that takes advantage of both self-fusion and deep neural networks. To boost the computational efficiency, we substitute self-fusion with a network that maps HN images to self-fusion of LN, which we call a `pseudo-modality'. From this smooth modality, we can easily extract a robust edge map to serve as a prior instead of a loss function. To combine the useful features from different modalities, we introduce a pseudo-multimodal fusion network (PMFN). It serves as a blender that can `inpaint'~\cite{bertalmio2000image} the fine details from HN on the canvas of clean layers from the pseudo-modality. The contributions of our work are the following:

\begin{itemize}
\item [$\blacklozenge$] A deep network to mimic the self-fusion process, so that the self-fusion of LN image becomes accessible at test time. This further allows the processing time to be sharply reduced.
\item [$\blacklozenge$] A pseudo-modality that makes it possible to extract clean gradient maps from high noise B-scans and provide compensation of layers and vessels in the final denoising result.
\item [$\blacklozenge$] A pseudo-multimodal fusion network that combines desired features from different sources such that the contribution of each modality is adjustable.
\end{itemize}

\section{Methods}
\begin{figure}[t]
  \includegraphics[width=1\linewidth]{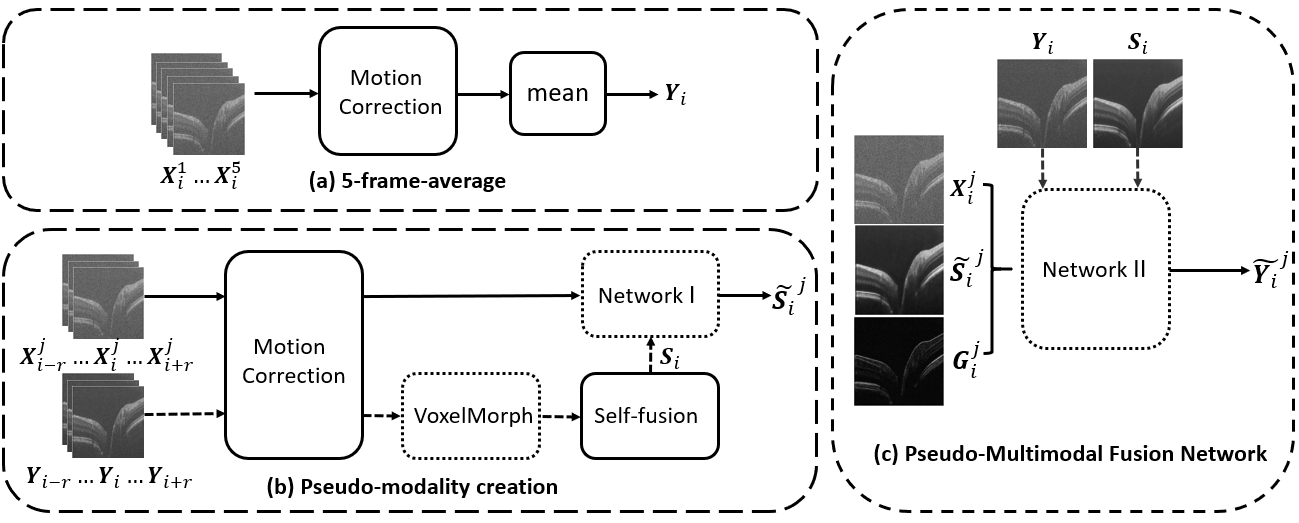}
  \caption{Processing pipeline. Dotted box refers to a deep learning network. Process on dash arrow exists only in training. Solid arrows are for both training and testing.}
  \label{fig:pipeline}
\end{figure}

Fig.~\ref{fig:pipeline} illustrates the overall processing pipeline.

\underline{\bf Preprocessing.} We crop every B-scan to size $[512,500]$ to discard the massive background that is not of interest. Then we zero-pad the image to $[512,512]$ for convenience in downsampling. 

\underline{\bf 5-frame average.} In our supervised learning problem, the ground truth is approximated by the low noise 5-frame-average B-scan (LN). The repeated frames at location $i$ are denoted by [$\bm{X}^1_i$, ..., $\bm{X}^5_i$] in Fig.~\ref{fig:pipeline}-a. Because of eye movement during imaging, some drifting exists between both repeated frames and adjacent B-scans. We apply a rigid registration for motion correction prior to averaging. 

\underline{\bf Pseudo-modality creation.}
For self-fusion, we need deformable registration between adjacent slices. This is realized by VoxelMorph~\cite{balakrishnan2019voxelmorph}, a deep registration method that provides deformation field from moving image to target. This provides considerable speedup compared to traditional registration algorithms. However, even without classical registration, self-fusion is still time-consuming. To further reduce the processing time, we introduce Network 1 to directly learn the self-fusion output. Time consumed by generating a self-fusion image of a B-scan drops from $7.303\pm 0.322$s to $0.253\pm 0.005$s. The idea allows us to also improve the quality of our pseudo-modality, by using $\bm{S}_i$, the self-fusion of LN $\bm{Y}_i$ images rather than that of HN images. Thus, Network \RNum{1} maps a stack of consecutive HN B-scans to self-fusion of LN.

In Fig.~\ref{fig:pipeline}-b, the noisy B-scan and its neighbors within a radius are denoted as {[$\bm{X}^j_{i-r}$, ..., $\bm{X}^j_{i+r}$]}, where $j=1,2,\dots,5$ represent the repeated frames. Their corresponding LN counterparts are named similarly, {[$\bm{Y}_{i-r}$, ..., $\bm{Y}_{i+r}$]}. The ground truth of Network \RNum{1} (i.e., the self-fusion of $\bm{Y}_i$) and its prediction are annotated as $\bm{S}_i$ and $\Tilde{\bm{S}_i}^j$ respectively. Since $\Tilde{\bm{S}_i}^j$ contains little noise, we can use its image gradient {$\bm{G}^j_i$}, computed simply via 3x3 Sobel kernels, as the edge map. 
 
 \underline{\bf Psudo-multimodal fusion network (PMFN).}  Fig.~\ref{fig:pipeline}-c shows the PMFN that takes a three-channel input. The noisy B-scan $\bm{X}^j_i$ has fine details including small vessels and texture, while the speckle noise is too strong to clearly reveal layer structures. The pseudo-modality $\Tilde{\bm{S}_i}^j$ has well-suppressed speckle noise and clean layers, but many of the subtle features are lost. So, merging the essential features from these mutually complementary modalities is our goal. To produce an output that inherit features from two sources, Network \RNum{2} takes feedback from the ground truth of both modalities in seeking for a balance between them. We use L1 loss for $\bm{Y}_i$ to punish loss of finer features and mean squared error (MSE) for $\bm{S}_i$ to encourage some blur effect in layers. The weight of these loss functions are determined by hyper-parameters. The overall loss function is:
\begin{equation}
    Loss=\alpha \sum_{x,y} |\Tilde{\bm{Y}_i}^j(x,y)-\bm{Y}_i(x,y)|+\frac{\beta}{N} \sum_{x,y} (\Tilde{\bm{Y}_i}^j(x,y)-\bm{S}_i(x,y))^2
\end{equation}
$N$ is the number of pixel in the image. Parameters $\alpha$ and $\beta$ are the weights of the two loss functions, and they can be tuned to reach a tradeoff between layers from the pseudo-modality and the small vessels from the HN B-scan. 

\section{Experiments}

\subsection{Data set}
OCT volumes from the fovea and optic nerve head (ONH) of a single human retina were obtained. For each region, we have two volumes acquired at three different noise levels (SNR=92dB, 96dB, 101dB). Each raw volume ($[N_{Bscan},H,W]=[500,1024,500]$) contains 500 B-scans of $1024 \times 500$ voxels. For every B-scan, there are 5 repeated frames taken at the same position (2500 Bscans in total) so that a 5-frame-average can be used as low-noise `ground truth'. Since all these volumes are acquired from a single eye, to avoid information leakage, we denoise fovea volumes by training on ONH data, and vice versa. 

\begin{figure}[t]
    \centering
    \includegraphics[width=\linewidth]{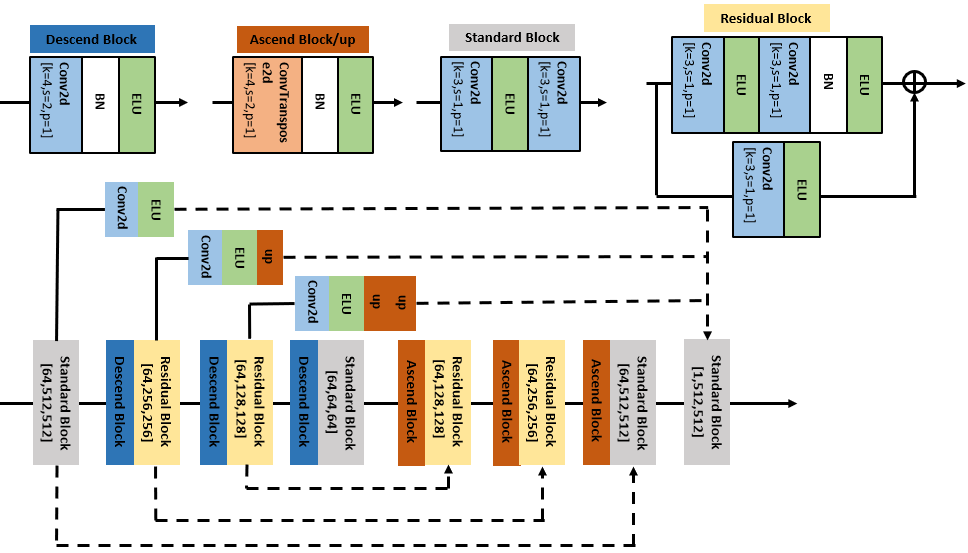}
    \caption{Network architecture. The solid line passes the computation result of the block while the dash line refers to channel concatenation. Arrays in main trunk blocks indicate the output dimension.}
    \label{fig:net}
\end{figure}

\subsection{Experimental design}
In this study, our goal is to show that the denoising result is improved by the processing pipeline that introduces the pseudo-modality. Thus, we will not focus on varying the network structure for better performance. Instead, we will use the Network \RNum{2} with single channel input $\bm{X}_i^j$ as the baseline. For this baseline, the loss function will only have feedback from $\bm{Y}_i$. We hypothesize that the relative results between single modality and pseudo-multimodal denoising will have a similar pattern for other architectures for Network \RNum{2}, but exploring this is beyond the scope of the current study.
Since the network architecture is not the focus of our study, we use the same multi-scale U-Net (MSUN) architecture, shown in Fig.~\ref{fig:net} and proposed by Devalla et al.~\cite{devalla2019deep}, for both Networks \RNum{1} and \RNum{2}.

The B-scan neighborhood radius for self-fusion was set at $r=7$. Among the five repeated frames at each location, we only use the first one ($\bm X_i^1$), except when computing the 5-average $\bm Y_i$. All the models are trained on NVIDIA RTX 2080TI 11GB GPU for 15 epochs with batch size of 1. Parameters in network are optimized by Adam optimizer with starting learning rate $10^{-4}$ and a decay factor of $0.3$ for every epoch. In Network \RNum{2}, we use $\alpha=1$ and $\beta=1.2$.

\section{Results}
\subsection{Visual Analysis}
 We first analyze the layer separation and vessel visibility in the denoised results.

\begin{figure}[t]
\centering
\begin{tabular}{cccc}
  & LN & MSUN & PMFN \\
 \rotatebox{90}{\hspace{0.4cm}SNR=92dB} & 
 {\includegraphics[width=0.31\linewidth,trim={0cm 0cm 0cm 11cm},clip]{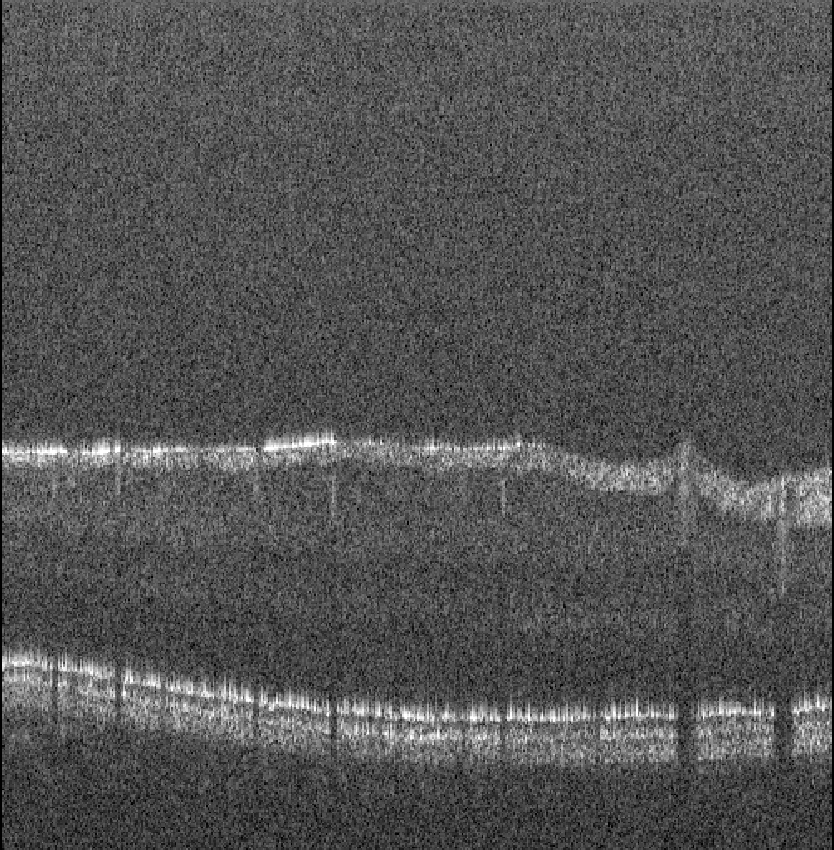}} &
 {\includegraphics[width=0.31\linewidth,trim={0cm 0cm 0cm 11cm},clip]{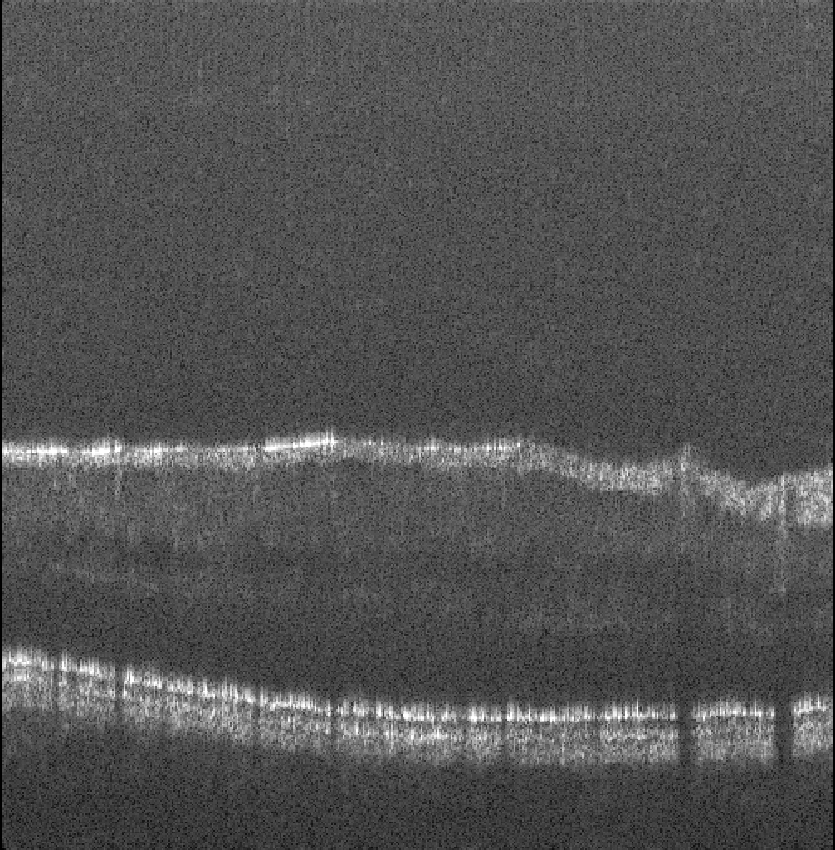}} &
 {\includegraphics[width=0.31\linewidth,trim={0cm 0cm 0cm 11cm},clip]{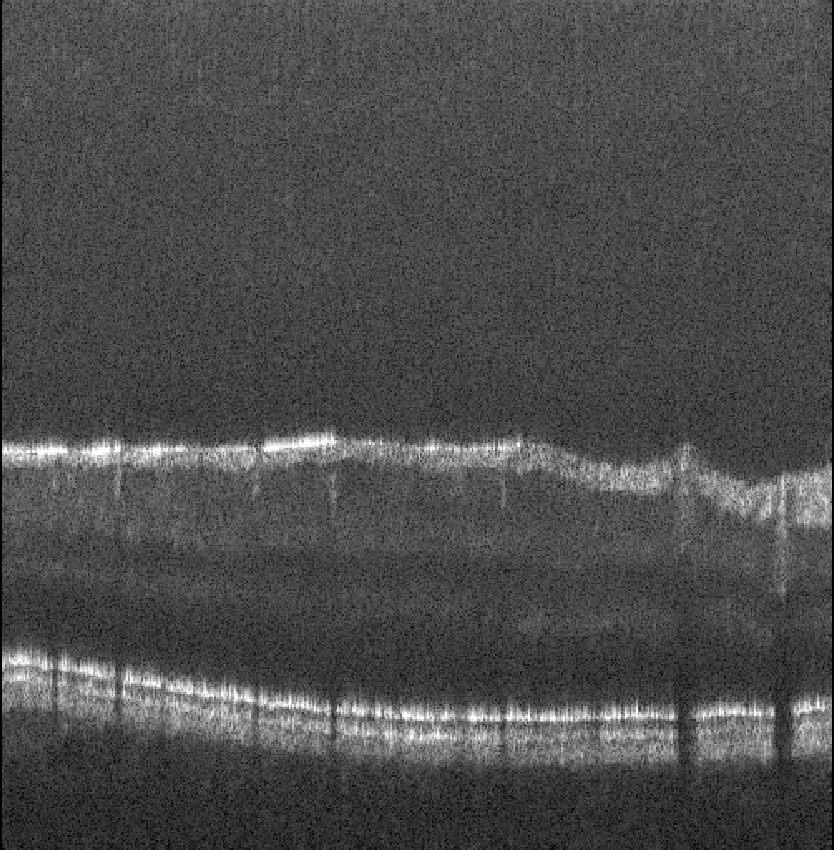}}  \\
 \rotatebox{90}{\hspace{0.3cm}SNR=96dB}& 
 {\includegraphics[width=0.31\linewidth,trim={0cm 2cm 0cm 7cm},clip]{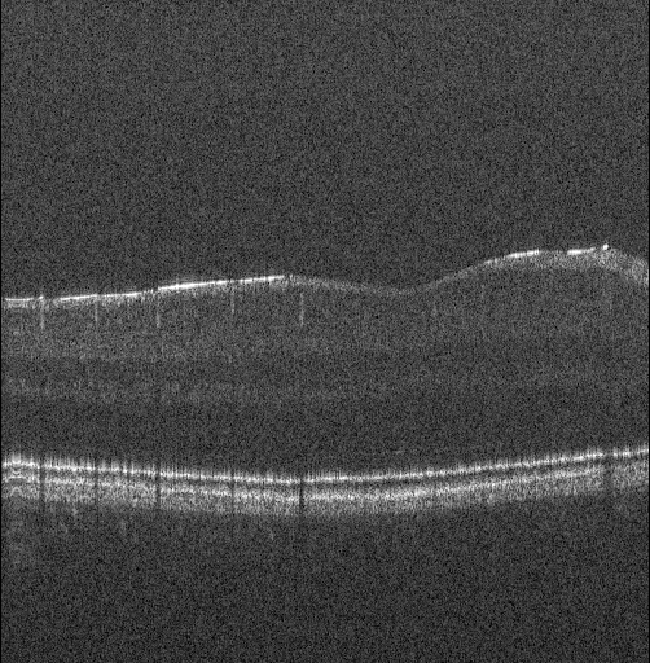}} &
 {\includegraphics[width=0.31\linewidth,trim={0cm 2cm 0cm 7cm},clip]{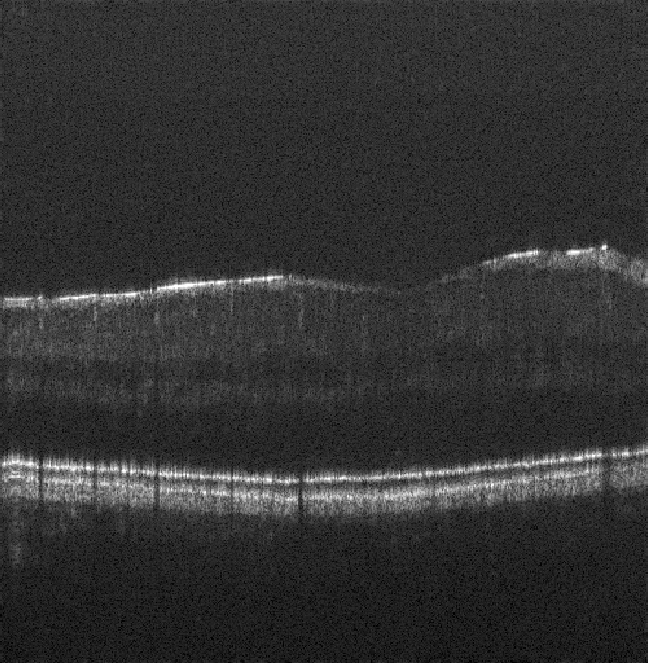}} &
 {\includegraphics[width=0.31\linewidth,trim={0cm 2cm 0cm 7cm},clip]{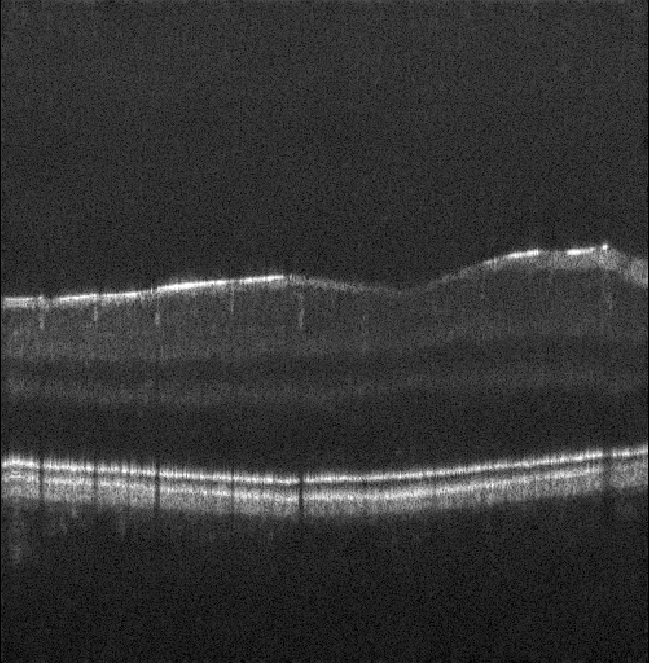}}  \\
 \rotatebox{90}{\hspace{0.3cm}SNR=101dB}& 
 {\includegraphics[width=0.31\linewidth,trim={0cm 0cm 0cm 12cm},clip]{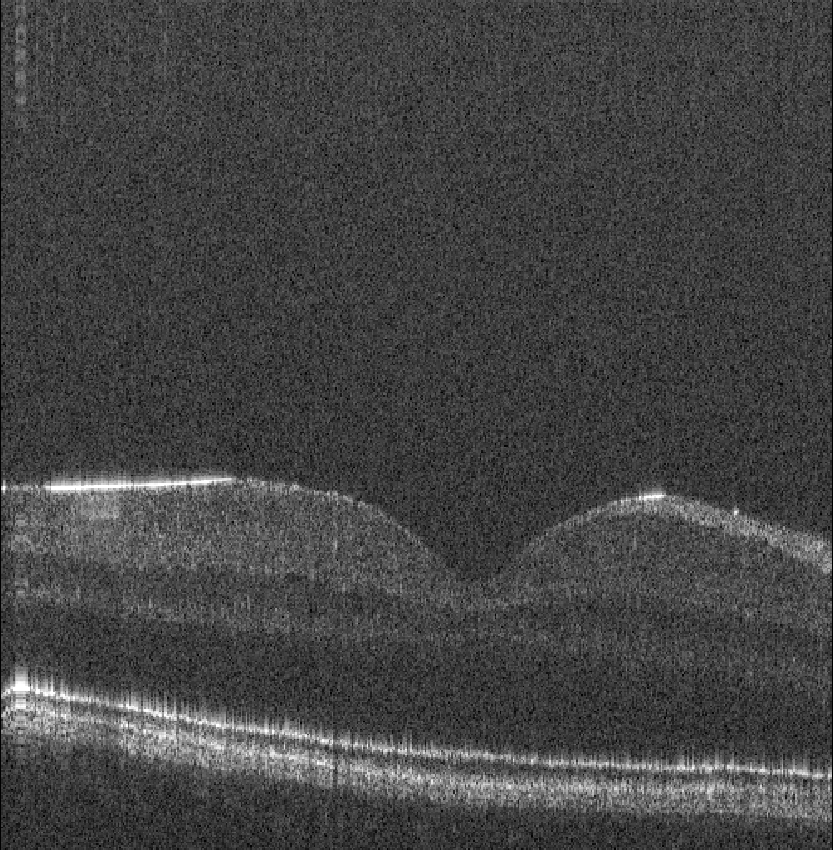}} &
 {\includegraphics[width=0.31\linewidth,trim={0cm 0cm 0cm 12cm},clip]{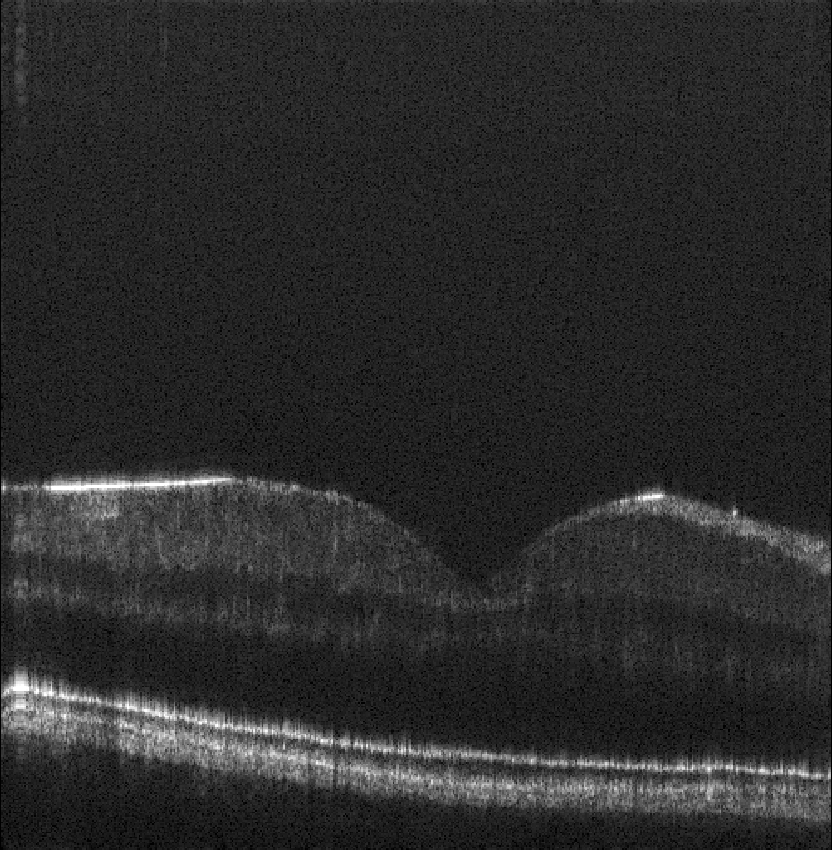}} &
 {\includegraphics[width=0.31\linewidth,trim={0cm 0cm 0cm 12cm},clip]{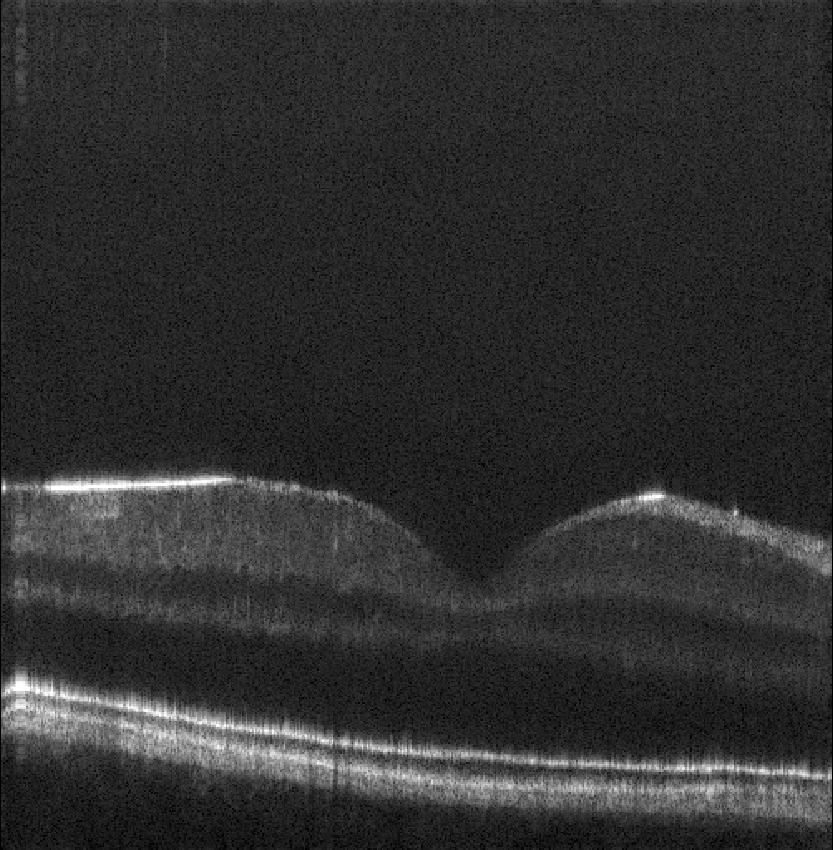}} 
\end{tabular}
\caption{Fovea denoising results for different input SNR. (Excess background trimmed.)}
\label{fig:result}
\end{figure}

Fig.~\ref{fig:result} displays the denoising performance of the proposed algorithm for different input SNR levels. Compared to the baseline model, we observe that PMFN has better separation between GCL and IPL, which enables the vessels in GCL to better stand out from noise. Moreover, the improvement of smoothness and homogeneity in outer plexiform layer (OPL) makes it look more solid and its border more continuous. In addition, the retinal pigment epithelium (RPE) appears to be more crisp. 

\begin{figure}[t]
\centering
    \begin{tabular}{ccc}
    \includegraphics[width=0.3\linewidth]{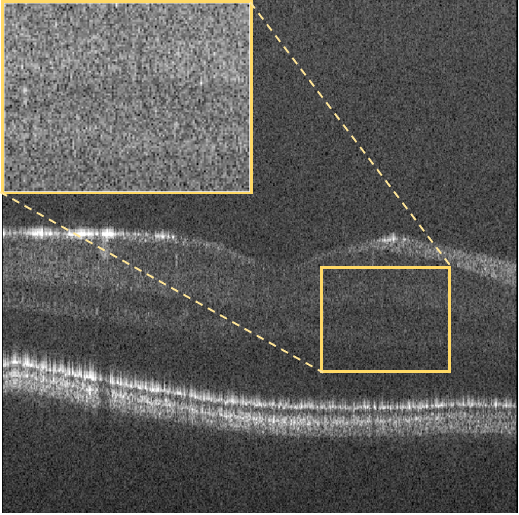}&
    \includegraphics[width=0.3\linewidth]{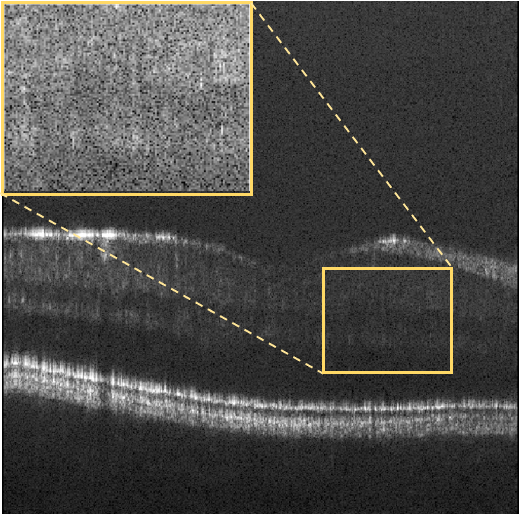}&
    \includegraphics[width=0.3\linewidth]{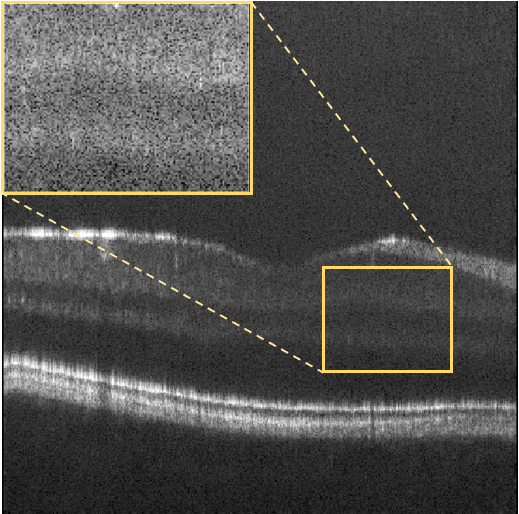}\\
    (a) LN &
    (b) MSUN &
    (c) PMFN \\
    \end{tabular}
    \\
    \begin{tabular}{cc}
    \includegraphics[width=.45\linewidth]{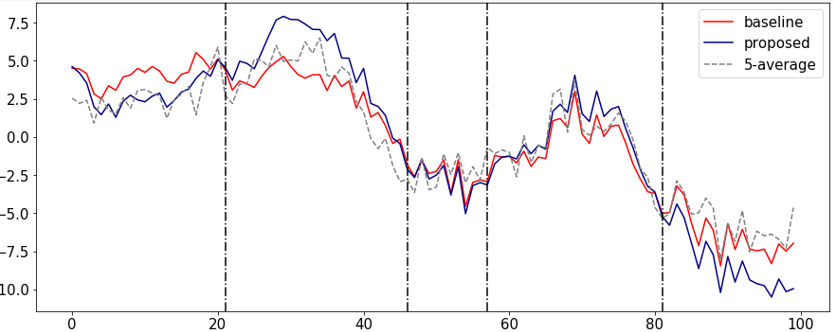}&
    \includegraphics[width=.45\linewidth]{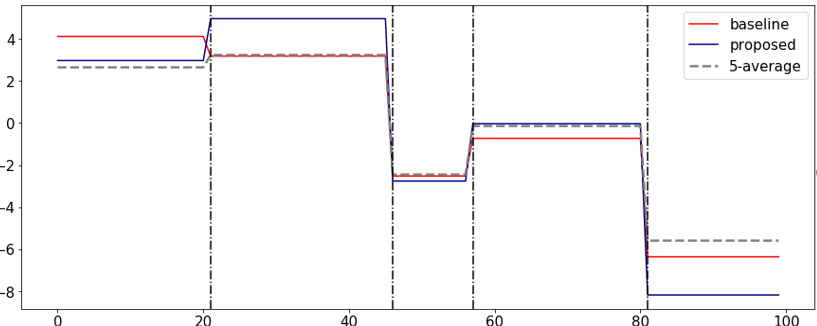}\\
    (d) Mean column intensity &
    (e) Mean layer intensity 
  \end{tabular}
\caption{Layer separation analysis. The top row shows an ROI containing 5 layers of tissue (GCL, IPL, INL, OPL, ONL) for each of (a) 5-average LN image, (b) baseline result and (c) PMFN result. (d) plots the intensity across the 5 layers within the ROI. (e) plots the mean intensity per layer. Vertical dashed lines approximate layer boundaries.}
\label{fig:layers}
\end{figure}

 In Fig.~\ref{fig:layers}, to better assess the layer separation, we focus on a B-scan with high speckle noise (SNR=92) that severely obscures the boundary between layers. In the top row, we zoom into a region of interest (ROI) that contains 5 tissue layers (from top to bottom): GCL, IPL, inner nuclear layer (INL), OPL and outer nuclear layer (ONL). As the baseline model learns only from the high noise B-scan, layer boundaries are not clear: GCL and IPL are indistinguishable, and although the INL and OPL are preserved, they are not as homogeneous as in the PMFN result. PMFN remedies these problems.
 
Another way of assessing the separability of layers or, in other words, the contrast between adjacent layers, is plotting the column intensity (Fig.~\ref{fig:layers}-d). Since the layers within the ROI are approximately flat, we take the mean vector along the row. In order to rule out the potential difference of intensity level, we normalize the mean vector with the average intensity of ROI.
\begin{equation}
    \bm{\Bar{v}}=\frac{1}{W}\sum_{i}^{W}\bm{v}_{i}-\bm{\mu}_{ROI}
\end{equation}
where W is the width of the ROI, $\bm{v}_{i}$ is a column vector in the window and $\bm{\mu}_{ROI}$ is a vector that has the mean of the ROI as all its elements. We plot the $\bm{\Bar{v}}$ for Fig.~\ref{fig:layers}-a, Fig.~\ref{fig:layers}-b and Fig.~\ref{fig:layers}-c in Fig.~\ref{fig:layers}-d. The border between layers are approximated with vertical dash lines for this visualization. In Fig.~\ref{fig:layers}-d, the proposed method tends to have lower intensity in dark bands and higher intensity in bright ones. This indicates that it has better contrast between adjacent layers. Fig.~\ref{fig:layers}-e
summarizes the mean intensity within each layer. Because of high intensity speckle noise, the baseline result completely misses the GCL-IPL distinction, whereas our method provides good separation.    

\subsection{Quantitative evaluation}
\begin{figure}[b]
\centering
    \begin{tabular}{ccc}
    \includegraphics[width=0.3\linewidth,trim={0cm 3cm 0cm 2cm},clip]{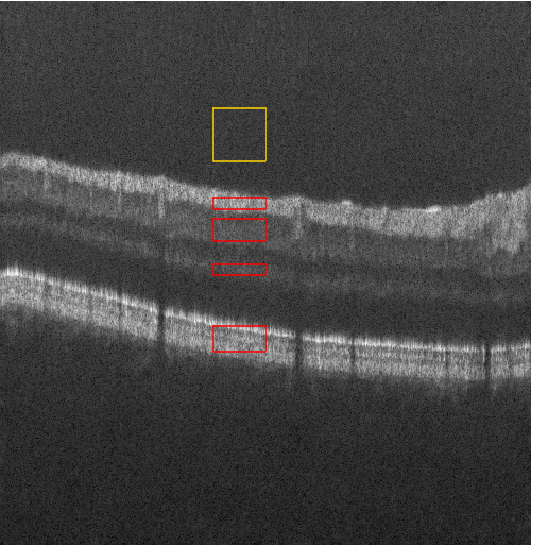}
    \includegraphics[width=0.3\linewidth,trim={0cm 3cm 0cm 2cm},clip]{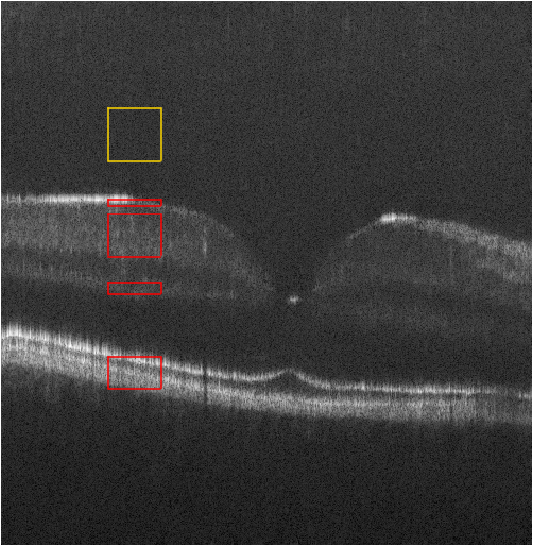}
    \includegraphics[width=0.3\linewidth,trim={0cm 3cm 0cm 2cm},clip]{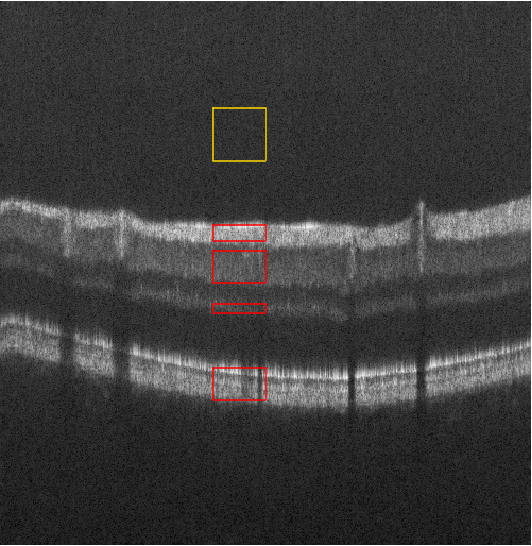}
    \end{tabular}
    \caption{Sample B-scans showing background (yellow) and foreground (red) ROIs used for SNR, CNR and PSNR estimation. 10 B-scans are chosen throughout the fovea volume to avoid bias.}
    \label{fig:snr}
\end{figure}

We report the signal-to-noise ratio (SNR), peak signal-to-noise ratio (PSNR), contrast-to-noise ratio (CNR) and structural similarity (SSIM) of our results. Normally, these metrics need an ideal ground truth without noise as a reference image. But such a ground truth is not available in our task, since the 5-frame-average LN image is far from being noiseless. Therefore, we make some adjustments to the original definitions of SNR and PSNR. We use
$
    SNR = 10 \log_{10}\left[\frac{\sum_{x,y}[f(x,y)]^2}{\sum_{x,y}[b(x,y)]^2}\right]
$
where $f(x,y)$ is the pixel intensity in foreground window and $b(x,y)$ is background pixel intensity. This assumes there is nothing but pure speckle noise in the background, and that the foreground window only contains signal. Similarly, the PSNR can be approximated by $
    PSNR = 10 \log_{10}\left[\frac{n_x n_y max[f(x,y)]^2}{\sum_{x,y}[b(x,y)]^2}\right]
$.
The $n_x$ and $n_y$ are the width and height of the ROI, respectively. Finally, the CNR is estimated by $
    CNR = \frac{|\mu_f-\mu_b|}{\sqrt{0.5(\sigma_f^2+\sigma_b^2)}}
$
where $\mu_f$ and $\sigma_f$ are the mean and standard deviation of the foreground region; $\mu_b$ and $\sigma_b$ are those of the background region. 

\begin{figure}[t]
\centering
    \begin{tabular}{cc}
    \includegraphics[width=.475\linewidth]{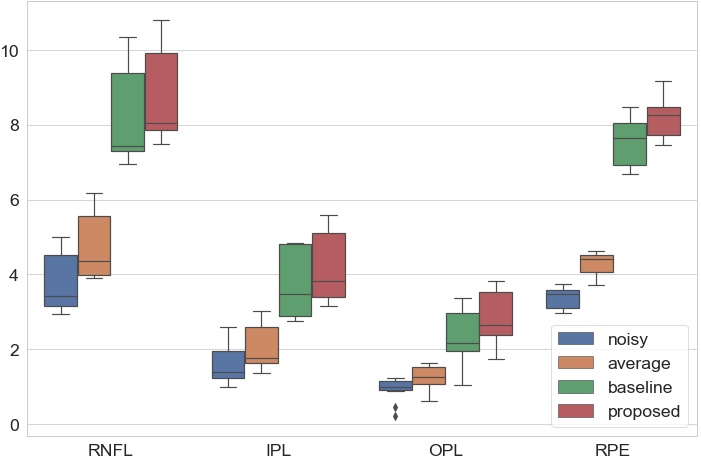}&
    \includegraphics[width=.475\linewidth]{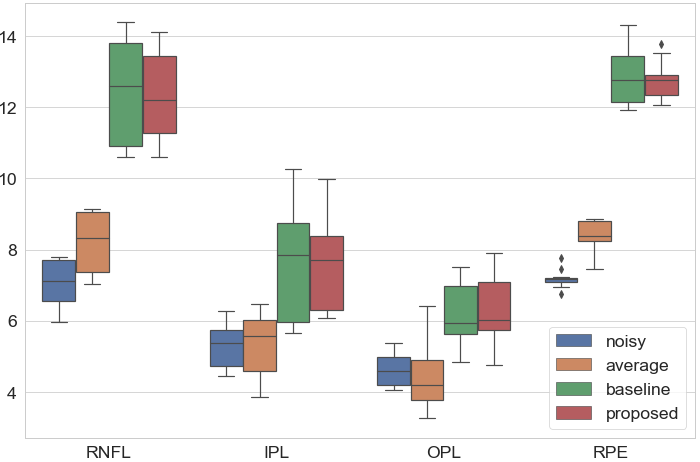}\\
    (a) SNR of each layer&
    (b) PSNR of each layer\\
    \includegraphics[width=.475\linewidth]{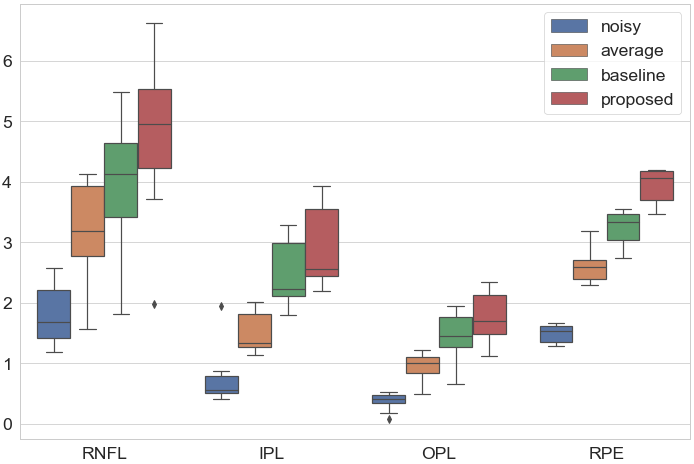}&
    \includegraphics[width=.475\linewidth]{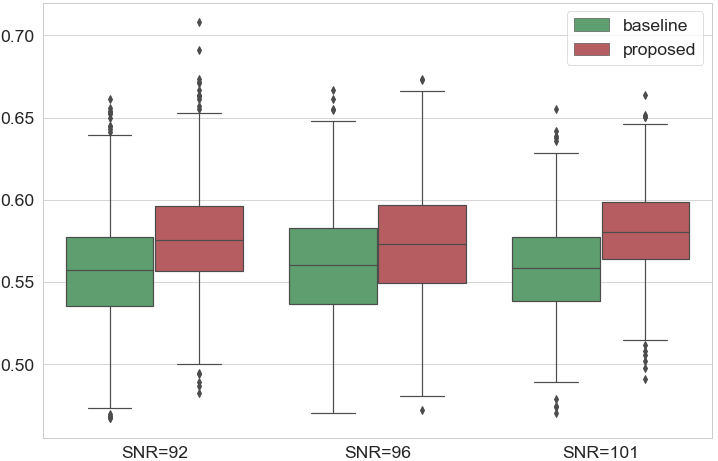} \\
    (c) CNR of each layer &
    (d) SSIM for input of different noise level
    \end{tabular}
    \caption{Quantitative evaluation of denoising results.}
    \label{fig:eval}
\end{figure}

Every layer has a different intensity level, so we report each metric separately for RNFL, IPL, OPL and RPE. We manually picked foreground and background ROIs from each layer, as shown in Fig.~\ref{fig:snr}, for 10 B-scans. To avoid local bias, these chosen slices are far apart to be representative of the whole volume. When computing metrics for a given layer, the background ROI (yellow box) is cropped as needed to match the area of the foreground ROI (red box) for that layer. 
Fig.~\ref{fig:eval} \textbf{(a)} to \textbf{(c)} display the evaluation result for SNR, PSNR and CNR respectively. For all layers, the proposed PMFN model gives the best SNR and CNR results, while the PSNR stays similar with the baseline multi-scale UNet model.

We also report the structural similarity index measure (SSIM)~\cite{zhou2004image} of the whole B-scan. The SSIM for each input SNR level is reported in Fig.~\ref{fig:eval}-d. The proposed method outperforms the baseline model for all input SNR.

\section{Conclusion and future work}
Our study shows that the self-fusion pseudo-modality can provide major contributions to OCT denoising by emphasizing tissue layers in the retina. The fusion network allows the vessels, texture and other fine details to be preserved while enhancing the layers. Although the inherent high dimensionality of the deep network has sufficient complexity, more constraints in the form of additional information channels are able to help the model converge to a desired domain. 

It is difficult to thoroughly evaluate denoising results when no ideal reference image is available. Exploring other evaluation methods remains as future work. Additionally, application of our method to other medical image modalities such as ultrasound images is also a possible future research direction.
\paragraph{\bf Acknowledgements.} This work is supported by Vanderbilt University Discovery Grant Program.

%
%
\bibliographystyle{splncs04}
\bibliography{mybibliography}
\end{document}